# Shutdown Margin for High Conversion BWRs Operating in Th-$^{233}$U Fuel Cycle

Y. Shaposhnik$^{a,*}$, E. Shwageraus$^{a}$, and E. Elias$^{b}$

$^{a}$ *Department of Nuclear Engineering*
*Ben-Gurion University of the Negev*
*POB 653, Beer Sheva 84105, Israel*

$^{b}$ *Faculty of Mechanical Engineering*
*Technion - Israel Institute of Technology*
*Technion city, 32000 Haifa, Israel*

**Abstract**

Several reactivity control system design options are explored in order to satisfy shutdown margin (SDM) requirements in a high conversion BWRs operating in Th-$^{233}$U fuel cycle (Th-RBWR). The studied has an axially heterogeneous fuel assembly structure with a single fissile zone "sandwiched" between two fertile blanket zones. The utilization of an originally suggested RBWR Y-shape control rod in Th-RBWR is shown to be insufficient for maintaining adequate SDM to balance the high negative reactivity feedbacks, while maintaining fuel breeding potential, core power rating, and minimum Critical Power Ratio (CPR). Instead, an alternative assembly design, also relying on heterogeneous fuel zoning, is proposed for achieving fissile inventory ratio (FIR) above unity, adequate SDM and meeting minimum CPR limit at thermal core output matching the ABWR power.

The new concept was modeled as a single 3-dimensional fuel assembly having reflective radial boundaries, using the BGCore system, which consists of the MCNP code coupled with fuel depletion and thermo-hydraulic feedback modules.

## 1. INTRODUCTION

The feasibility of thorium utilization in BWRs based on Th-$^{233}$U fuel cycle has been recently investigated in numerous studies [Shaposhnik et al. 2013; Ganda et al., 2011; Kim and Downar, 2002; Todosow et al., 2005; Takaki et al., 2007; Volaski et al., 2009; Raitses and Todosow, 2010] motivated by the need to design a viable self-sustainable, with respect to fissile material requirements, light water cooled reactors, which could be an alternative to the more complex and costly fast reactor technology. Shaposhnik et al. (2013) have presented an optimized fuel assembly configuration with a single axial fissile zone situated between two fertile blanket zones that enabled a self-sustainable fuel cycle based on Th-$^{233}$U mixed oxide fuel (Th-RBWR). The optimization addressed fuel temperature and Critical Power Ratio (CPR) limits as well as selection of the optimal dimensions of the fissile and fertile blanket zones. However, in order to meet CPR margin, the core rated power had to be reduced by about 30% compared to standard ABWR core power (3973 MW$_{th}$). Furthermore, the high magnitude of reactivity feedbacks (Doppler and void coefficients) makes it difficult to achieve adequate shutdown margin.

The objective of the present study is to extend the previous design in order to achieve adequate shutdown margin, while matching the standard ABWR core power and maintaining Fissile Inventory Ration (FIR) of the discharge fuel above unity. This is carried out by examining a wide range of design options and applying a number of limiting constraints.

## 2. METHODOLOGY

In order to achieve high breeding ratio, fertile and fissile nuclides should be spatially separated to reduce their competition for neutron absorption and thus maximizing capture rate in fertile nuclides. An optimal configuration and composition of the fissile and fertile fuel zones with high breeding ratio were studied by Shaposhnik et al. (2013). This design consisted of a tightly packed hexagonal fuel lattice operating at high core average void fraction (60% versus about 40% in a typical BWR) in order to harden the neutron spectrum and improve breeding. The reactivity coefficients (Doppler and coolant void) were found to be negative during the entire irradiation period. However, the high magnitude of both reactivity feedbacks made it difficult to achieve adequate shutdown margin. Thorium fuel is known to have about twice as negative Doppler coefficient (DC) as a typical $UO_2$ fuel. The magnitude of the void coefficient (VC) in the High Conversion RBWR core is about the same as in a standard BWR. However, at nominal Hot Full Power (HFP) the core has much higher void fraction, while at cold shutdown state, Cold Zero Power (CZP), the void fraction is reduced to zero. Therefore, the total reactivity change between HFP and CZP states is significantly increased compared to standard BWR. At such conditions, it may be impractical to compensate for the excess reactivity with control and safety rods proposed in the original RBWR-AC design [Feng et al., 2011]. Moreover, at CZP, the neutron leakage from the seed to the blanket zone generally decreases, leading to further increase in reactivity.

In the following, the shutdown margins issue is considered by extending the Th-RBWR core proposed in Shaposhnik et al. (2013) to address three design questions:

- Can the design satisfy shutdown margin using the Y-shaped control rod (CR) configuration (as in the RBWR-AC design [Feng et al., 2011])?,

- What control material and control rods configuration assure the shutdown margin?

- How the reactivity control and shutdown margin requirements affect the breeding performance and achievable power of the core?

## 2.1. Definitions

Before proceeding with the description of the computational procedure in details, we summarize the terminology used throughout the paper:

FIR – Fissile Inventory Ratio is defined as the core fissile inventory at a specific time divided by the initial core fissile inventory. FIR was calculated by summing up the masses of fissile nuclides ($^{233}$U and $^{235}$U) at time t, with an addition of $^{233}$Pa. The reason for counting $^{233}$Pa as a fissile nuclide is that, in case of core shutdown, all $^{233}$Pa relatively quickly decay into fissile $^{233}$U.

HFP – Hot Full Power is taken as 7.1 MPa steam dome pressure with typical void distributions and fuel temperature ranging from 600 to 900 K according to axial power distribution in the assembly at nominal power rating.

HZP – Hot Zero Power refers to pressure of 7.1 MPa with fuel and coolant temperatures both at 560 K.

CZP – Cold Zero Power refers to 0.1 MPa pressure with fuel and coolant temperatures at 320 K.

SDM – Shutdown Margin is defined as the amount of reactivity by which a full reactor core is subcritical at its present condition assuming all full-length control blades or rod cluster assemblies (shutdown and control) are fully inserted.

## 2.2. Computational tools

The analyses were performed using BGCore code system [Fridman et al., 2008a; Kotlyar et al., 2011], which couples the Monte Carlo code, MCNP, with fuel depletion and thermal hydraulic feedback modules.

The system uses a multi-group approach for calculating one-group cross sections required for the depletion calculations [Fridman et al., 2008b] and provides significant calculation speedup. Thermal feedback is based on a sub-channel analysis code (THERMO) capable of handling single and two-phase flow [Shaposhnik et al., 2012]. At every depletion time step, it provides fuel temperature as well as coolant temperature and density distributions, which are automatically updated in an iterative manner until desired convergence is achieved [Kotlyar et al., 2011]. The accuracy and validity of the thermal hydraulic module was verified against results of other codes [Shaposhnik et al., 2008; Kotlyar et al., 2011], such as the ELOCS system and DYN3D.

Temperature dependent cross section libraries based on JEFF3.1 [Koning et al., 2006] evaluated data were used.

In order to assure accuracy of the results, Shannon entropy [Ueki et al., 2003] was used as a measure of the fission source distribution convergence. Since monitoring the statistical error in the $k_{eff}$ eigenvalue alone does not necessarily ensure the convergence.

**2.3. Assembly geometry with Y-shape control rod**

In the reference Th-RBWR core calculations [Shaposhnik et al. (2013)], the simplified fuel assembly model assumed equal assembly lattice pitch in all directions (i.e., the water gaps were assumed to be of the same size on all sides of the assembly). As suggested by Feng et al., 2011, the control blade follower materials and the gap water densities were adjusted to match the reactivity and absorption reaction rates of the model with real detailed geometry. This simplified geometry was found to be adequate for evaluating the equilibrium cycle core composition when all control rods are fully withdrawn (All Rods Out - ARO). However, in order to evaluate SDM, the Y-shape control rods located between the fuel assemblies had to be modeled in details. Figure 1 presents a typical Y-shape control rod design and composition as proposed by Hitachi, which is used in RBWR assemblies [Feng et al., 2011]. Figure 2 depicts the assembly layout with ARO and ARI in case of detailed geometry description of the Y-shape control blade as used in this study.

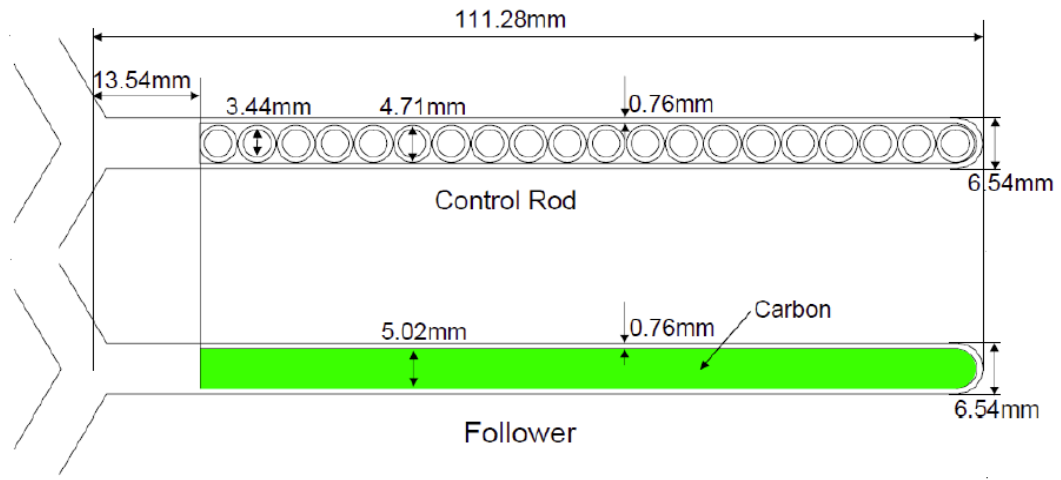

**Figure 1:** RBWR-AC Y-shape control rod geometry and composition [Feng et al., 2011]

| | | |
|---|---|---|
| **Detailed geometry (ARO)** | | |
| **Detailed geometry (ARI)** | | |

**Figure 2:** Y-shape control rod: detailed geometry models

The reactivity worth of all the control rods, $\rho_{allCRD}$, was calculated using multiplication factors of the assembly with all control rods withdrawn, $k_{ARO}$, and all control rods inserted, $k_{ARI}$ as given in Equation 1:

$$\rho_{allCRD} = \rho_{ARO} - \rho_{ARI} = \frac{k_{ARO}-1}{k_{ARO}} - \frac{k_{ARI}-1}{k_{ARI}} = \left(\frac{1}{k_{ARI}} - \frac{1}{k_{ARO}}\right) \qquad (Eq.\ 1)$$

## 3. RESULTS

### 3.1 Comparison of SDM in Th-RBWR and RBWR-AC with Y-shape control rod

As a first step, we examined the possibility of meeting the SDM requirements in the Th-RBWR design using the Y-shape control rod as in RBWR-AC design [Feng et al., 2011]. The RBWR-AC design was used as a reference case for SDM calculations since the radial dimensions of both assemblies, RBWR-AC and Th-RBWR, are identical.

The RBWR-AC design, described in Table 1 and Fig. 3, is able to achieve a breeding ratio of 1.01 (amount of fissile Pu in spent fuel divided by initial amount of fissile Pu), using (U,TRU)$O_2$ mixed oxide fuel, high coolant void fraction, and axially-alternating fissile and fertile fuel zones [Feng et al., 2011].

Table 1. RBWR-AC pin and assembly geometry

| *Parameter* | *Value* | *Units* |
|---|---|---|
| Fuel Diameter | 0.87 | [cm] |
| Gap Thickness | 0.075 | [cm] |
| Cladding Thickness | 0.06 | [cm] |
| Pin Diameter | 1.005 | [cm] |
| Pin Pitch | 1.135 | [cm] |
| Pitch/Diameter Ratio | 1.13 | |
| Assembly Inner Flat-to-Flat | 18.91 | [cm] |
| Assembly Can Thickness | 0.24 | [cm] |
| Assembly Pitch (narrow) | 19.47 | [cm] |
| Assembly Pitch (wide) | 19.92 | [cm] |

Figure 3 shows the axial fuel composition of the RBWR-AC assembly and Th-RBWR. The lower fissile zone in RBWR-AC design consists of (U,TRU)O$_2$ with fissile Pu making up 20.2 wt% of HM, whereas the upper fissile zone has 14.9 wt% fissile Pu.

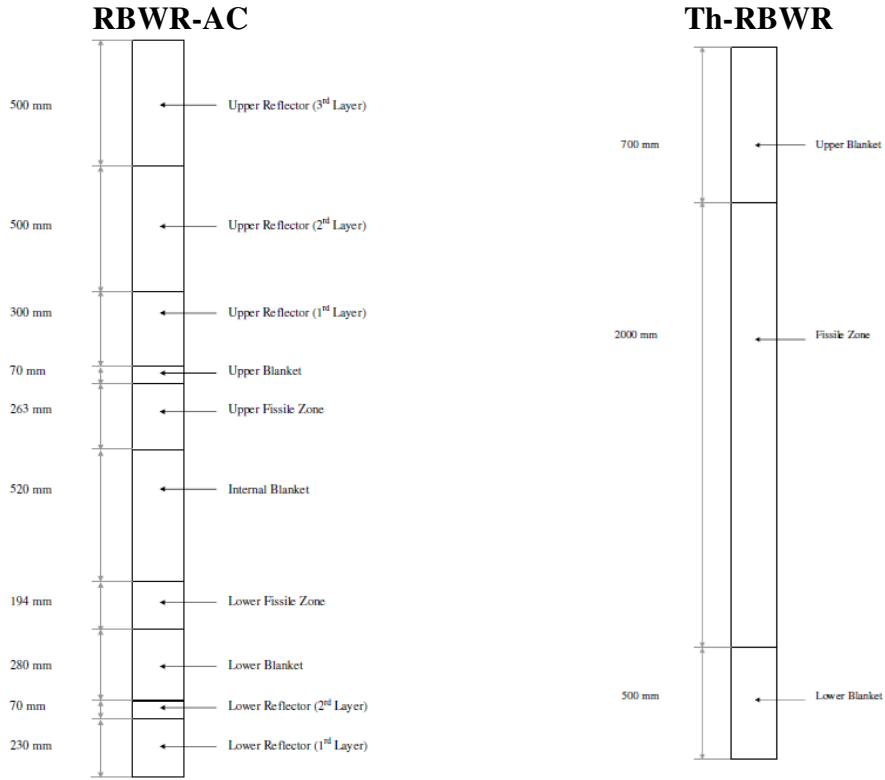

**Figure 3: RBWR-AC and Th-RBWR fuel assembly axial zones layout**

The control rod and follower geometries and compositions are shown in Figure 2. Note that the control material is 90% $^{10}$B enriched B$_4$C. Table 2 and Table 3 present the SDM at BOL for RBWR-AC and Th-RBWR assembly respectively. Each table presents the assembly lattice criticality values for HFP, HZP and CZP conditions. In a typical LWR, the SDM at EOL is much more problematic than at BOL, because MTC becomes progressively more negative with burnup. However, in the Th-RBWR design, both Doppler and coolant void reactivity coefficients are less negative at EOL compared to BOL [Shaposhnik et al., 2013]. Therefore, the SDM analysis was performed for the more restrictive BOL conditions.

The calculations were performed for a single assembly. Radial core leakage as well as the fact that a real core may contain a mix of partially burnt fuel batches was not taken into account at this stage. It is, therefore, expected that the core reactivity control requirements will be smaller than the values shown in these tables. Nevertheless, it was assumed that Hitachi RBWR design has adequate SDM and, therefore, we compared all our SDM values to the Hitachi ones.

**Table 2. SDM for RBWR-AC design at BOL**

| RBWR-AC | ARO | | ARI | | SDM |
|---|---|---|---|---|---|
| Case | k-inf | $\Delta$ k-inf | k-inf | $\Delta$ k-inf | [$\rho_{allCRD}$] |
| HFP | 1.05689 | 0.00032 | 0.98930 | 0.00029 | 0.065 |
| HZP | 1.06196 | 0.00028 | 0.99822 | 0.00032 | 0.060 |
| CZP | 1.07332 | 0.00029 | 1.01345 | 0.00032 | 0.055 |

**Table 3. SDM for Th-RBWR design at BOL**

| Th-RBWR | ARO | | ARI | | SDM |
|---|---|---|---|---|---|
| Case | k-inf | $\Delta$ k-inf | k-inf | $\Delta$ k-inf | [$\rho_{allCRD}$] |
| HFP | 1.05124 | 0.00027 | 0.87807 | 0.00022 | 0.188 |
| HZP | 1.42042 | 0.00021 | 1.24124 | 0.00030 | 0.102 |
| CZP | 1.51848 | 0.00025 | 1.35171 | 0.00027 | 0.081 |

It is noted that the reactivity values at BOL-HFP-ARO conditions are very close for both designs. However, at BOL-CZP-ARO the Th-RBWR assembly has excess reactivity of about 50,000 pcm compared to about 7,000 pcm in the RBWR-AC design. This can be attributed to a number of effects:

- Doppler reactivity coefficient is twice as high in Thorium as in Uranium fuels.
- Fissile properties of $^{233}$U are slightly more superior to Plutonium in the thermal spectrum.
- The leakage from fissile to fertile regions at CZP in a single axial seed zone assembly is smaller. One of the distinct features of RBWR-AC design is the axial configuration of the material zones (alternating fissile and fertile fuel regions). In RBWR-AC design, during operation at HFP, the axial fissile zones have an exchange of neutrons due to the high void fraction and thus long mean free path of neutrons. However, when the reactor is shutdown and the void fraction drops to zero, the neutrons mean free path reduces considerably and the two fissile zones become neutronically decoupled. This feature, in turn, increases the leakage of neutrons from the fissile zones at CZP and reduces the SDM requirements.

The results presented in Tables 2 and 3 suggest that the Y-shape control rod as used in RWBR-AC design will not be sufficient in the Th-RBWR design to meet the SDM requirements.

In order to address this challenge in Th-RBWR design, the control material, configuration of the CR, as well the fuel to control material mass (or volume) ratio are reexamined. This includes examining the possibility of:
- Using alternative absorbing materials in the Control Rods
- Optimization of axial power shape by axially varying fissile content
- Use of fuel integral burnable absorbers (IBA)
- Modification the fuel to control rod volume ratio (smaller assemblies)
- Use of Rod Cluster Control Assemblies (RCCA spider) similar to PWR's

**3.2 Study of alternative control materials**

A reactivity control material generally comprised of chemical elements with sufficiently high capture cross section to compensate for the excess reactivity during the fuel irradiation campaign or due to the changes in the reactor power or operating conditions. Classical control materials are boron, hafnium, gadolinium, combination of silver, indium and cadmium (Ag-In-Cd or AIC) and some others. In addition, the control material should be in a chemical form (compound), which is chemically stable and compatible with other core materials in the range of operating and off-normal temperatures and pressures. Furthermore, it should be preferably resistant to radiation damage and maintain its high neutron absorption properties for a reasonable time (e.g., at least for one operating cycle). Some of the common compounds include Ag-In-Cd metal alloy, boron carbide, hafnium diboride, gadolinium oxide, gadolinium diboride etc [IAEA 1995]. The choice of materials is also affected by the energy of neutrons in the reactor because some materials have high absorption cross section only in some specific energy range. For example, Cd and Gd have particularly high absorption cross section in the thermal region. While, Boron and Hafnium in addition to high thermal absorption cross section are relatively strong absorbers at resonance neutron energies as well, making them suitable for use in different reactor types or equally effective under different operating conditions. This is the logic behind combining Ag-In-Cd in the same alloy as used in most PWRs. Each element in the alloy has significant absorption cross section in a certain energy range but, all together covering the entire neutron energy spectrum.

In the following, five potential control material compounds are examined as candidates for Y-shape control rods; namely, $B_4C$, Ag-In-Cd (80/15/5$^w/_o$), $Gd_2O_3$, a mix of $Gd_2O_3$ (10$^w/_o$) + $B_4C$(90$^w/_o$) and $HfB_2$. These compounds are widely used in the nuclear industry for reactivity control [IAEA 1995] and also because their absorbing efficiency covers a wide range of the neutron energy spectra. For example, Gd dominates in the thermal range (<1eV) while Hf is superior in the epithermal range (1eV<E<150eV). In all cases, boron was 90% enriched with $^{10}B$ isotope.

Table 4 presents the results obtained for the reference Th-RBWR assembly at BOL and CZP operating conditions.

**Table 4. SDM for reference Th-RBWR case at BOL for CZP with several options for control materials used in the Y-shape control rod**

| Control materials used in the Y-shape Control Rod | ARO | | ARI | | SDM |
|---|---|---|---|---|---|
| | k-inf | $\Delta$ k-inf | k-inf | $\Delta$ k-inf | [$\rho_{allCRD}$] |
| $B_4C$ | 1.51848 | 0.00025 | 1.35171 | 0.00027 | **0.081** |
| Ag-In-Cd $_{(80\%/15\%/5\%)}$ | | | 1.44196 | 0.00029 | **0.035** |
| $Gd_2O_3$ | | | 1.44877 | 0.00029 | **0.032** |
| $Gd_2O_3$ $_{(10\%)}$ + $B_4C_{(90\%)}$ | | | 1.36496 | 0.00030 | **0.074** |
| $HfB_2$ | | | 1.36521 | 0.00026 | **0.074** |

Table 4 shows that a $B_4C$ loaded control rods has the highest negative reactivity worth despite the fact that Gadolinium (Gd) and Cadmium (Cd) have higher thermal absorption cross section as shown in Figure 4.

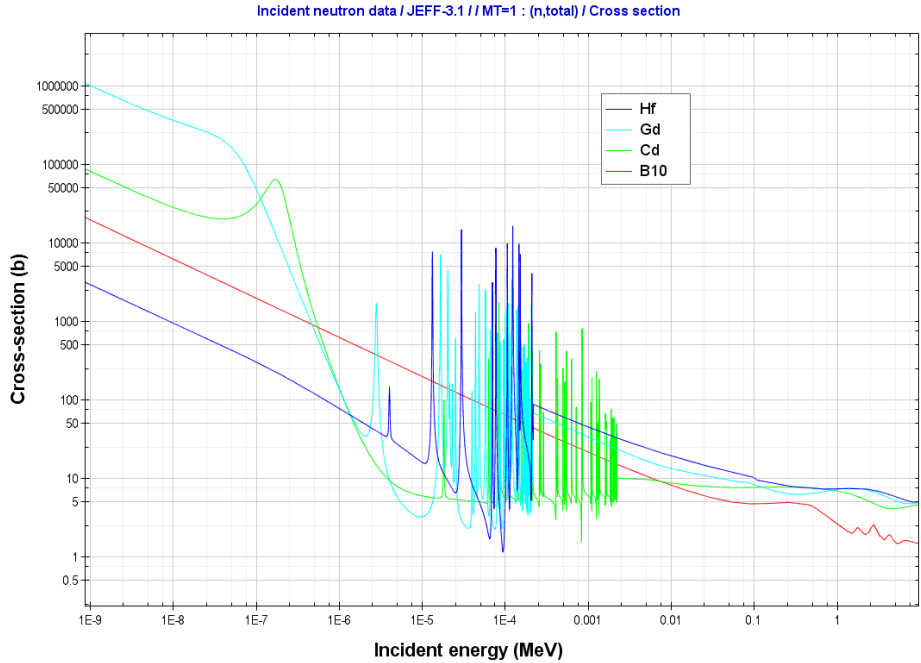

**Figure 4. Total neutron cross section of selected isotopes (JEFF3.1)**

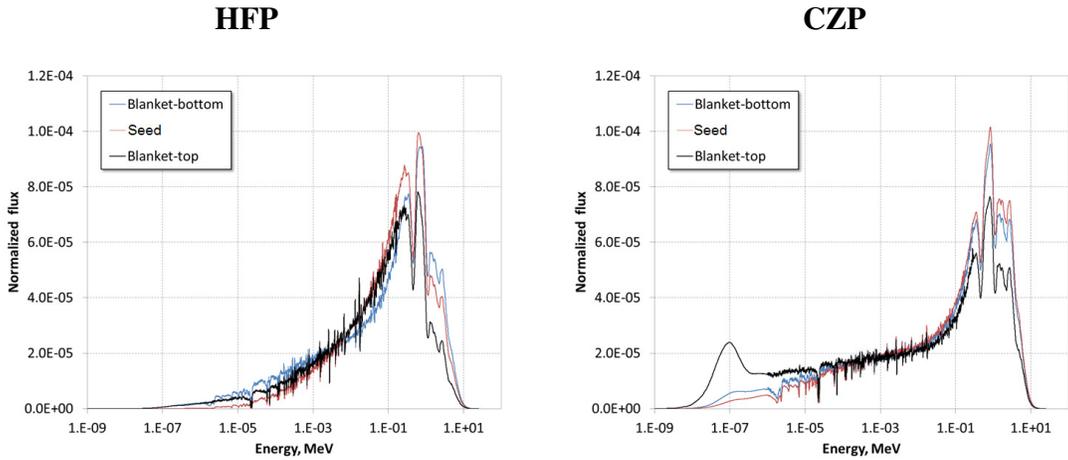

**Figure 5. Average neutron flux at BOL in Th-RBWR assembly at HFP and CZP**

The advantage of $B_4C$, in this case, is due to (1) higher atom density of the absorber in the control rod compound and (2) the neutron spectrum shape at CZP conditions (Fig. 5). The Th-RBWR assembly has a tightly packed hexagonal lattice, which significantly hardens the spectrum as presented in Figure 5. Therefore, the high thermal absorption cross section of Gd and Cd is not very effective. This effect is illustrated in Figure 6, which presents the assembly multiplication factor at BOL for CZP conditions with all rods in (ARI) as a function of Cd content in AIC control rod. The effectiveness of the control rods containing high fraction of Cadmium is actually reduced with an increase in Cd weight fraction because of the saturation of Cd absorption effect due to strong self-shielding and simultaneous reduction in Ag and In density, which are both epithermal absorbers. In the case of Th-RBWR assembly, the spectrum is harder than in a typical BWR or PWR, so that even at CZP conditions, the resonant absorbers play significant role.

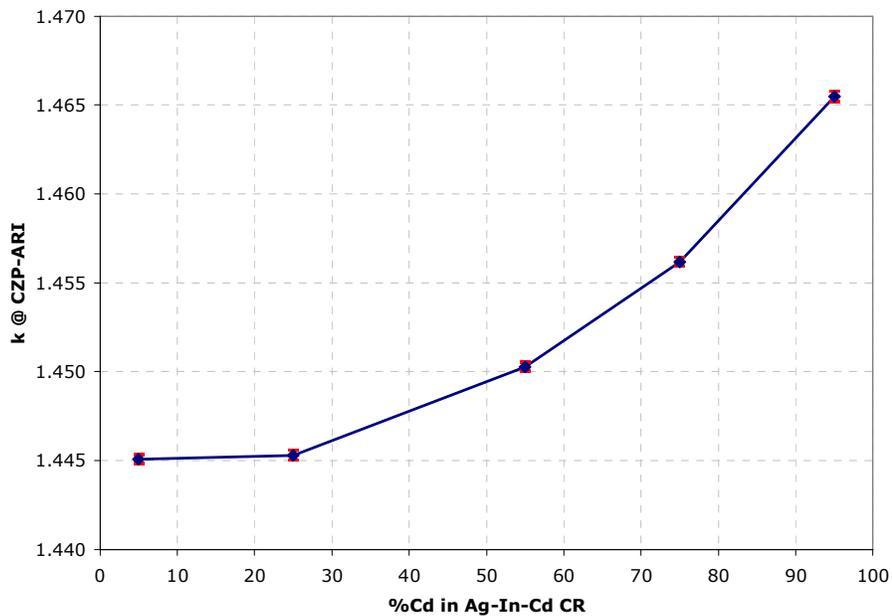

**Figure 6. Effect of Cadmium content in Ag-In-Cd control rods at BOL-CZP-ARI**

The main conclusion from this exercise is that $B_4C$ containing high fraction of $^{10}B$ is the most effective absorber material, closely followed by Hf. Therefore, we examined the possibility of replacing the Y-shape rod structural materials (Fig. 2) with Hf.

Table 5 presents the SDM for three control rod sheath configuration options:

Case I: The SS sheath (in the follower as well as in the absorber section) was replaced with Hf

Case II The SS sheath in the absorber section was replaced with Hf, while in the follower section; the graphite remained coated with the SS sheath.

Case III: The SS sheath in the absorber section was replaced with Hf and the whole follower section was replaced by a solid Zr.

**Table5. SDM for Th-RBWR assembly at BOL for CZP with several options for sheath material used in the Y-shape control rod**

| Control materials used in the Y-shape control rod | ARO | | ARI | | SDM |
|---|---|---|---|---|---|
| | k-inf | $\Delta$ k-inf | k-inf | $\Delta$ k-inf | [$\rho_{allCRD}$] |
| Case I | **1.13037** | *0.00026* | **1.07236** | *0.00026* | **0.048** |
| Case II | **1.51848** | *0.00025* | **1.07252** | *0.00026* | **0.274** |
| Case III | **1.52323** | *0.00023* | **1.07252** | *0.00026* | **0.276** |

As shown in Table 5, replacing the SS sheath with Hf sheath indeed improves the SDM considerably. Nevertheless, in Case I (where both the absorber and the follower sheath were replaced with Hf) it also affects the ARO conditions due to parasitic absorption in the follower section. This effect was eliminated in Case II, where only the absorption section of the sheath was replaced by Hf. In Case III we have used a solid Zircaloy follower instead of a SS coated Carbon. However, these designs may not be practical due to manufacturing constrains of the control rods, which are beyond the scope of this study. It may prove to be difficult to manufacture the control rod sheath composed of two different structural materials. It is, therefore, concluded that changing absorber materials in the Y-shape control rod is ineffective and insufficient for providing adequate shutdown margin. In the following we pursue other design options.

## 3.4 Variable axial enrichment

In order to reduce the peak linear heat generation rate, fissile content in the seed axial zone was varied (split into multiple enrichment zones). In the original design with nearly uniform axial enrichment distribution, the power peak occurs close to the interface between the seed and the bottom blanket. This power peak implies also high leakage of neutrons to the blanket and therefore larger contribution of leakage effect to already very negative void coefficient of reactivity. If axial enrichment is redistributed such that the power peak is moved away from the seed boundary, the magnitude of this negative contribution could be reduced. Thus, the void coefficient will become less negative reducing the SDM requirements. The seed zone was divided into five regions with variable fissile content (3.5-7.5$^{w}/_{o}$ of $^{233}$U), while adjusting the assembly average fissile content in order to match the fuel cycle length of the reference design. Figure 7 presents the assembly $k_{inf}$ and FIR as a function of time for the reference Th-RBWR design with uniform fissile content (6.5$^{w}/_{o}$) along the seed section (original design) and the design with axially varying fissile content.

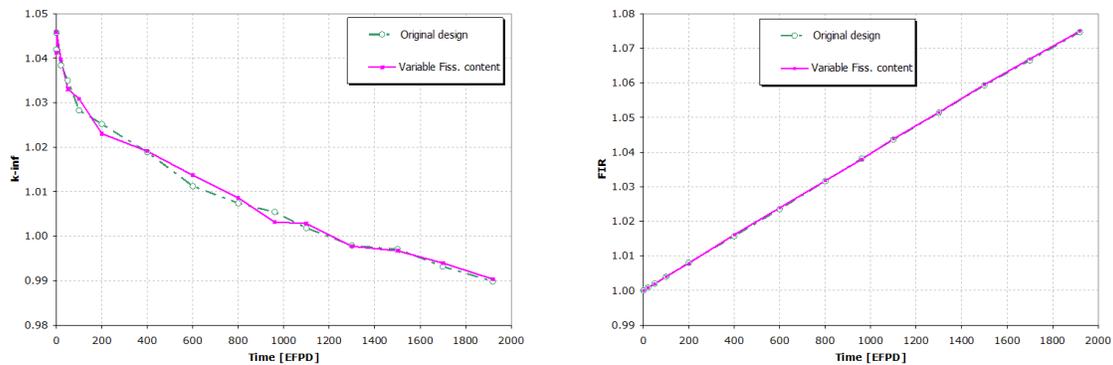

**Figure 7. $k_{inf}$ (left) and FIR (right) as a function of time for variable vs. uniform fissile content distribution in Th-RBWR assembly**

As observed from the figure, the transition to variable enrichment has practically no effect on either fuel cycle length or FIR. It also allows somewhat reducing the total initial fissile content (from 6.5$^{w}/_{o}$ to 6.0025 $^{w}/_{o}$) in the assembly. This reduction in fissile content is due to a lower leakage from the seed region to the blankets. As a result, the fissile content required to achieve a certain cycle length is also reduced. This, in turn, increases the conversion ratio in the seed, which is, in fact, not negligible and has significant contribution to the overall fissile material balance over the irradiation time. Therefore, the overall breeding performance is not affected despite the fact that the neutron leakage to the lower blanket is reduced.

Figure 8 presents the linear heat generation rate along the seed section in the Th-RBWR assembly at BOL for uniform and variable fissile content.

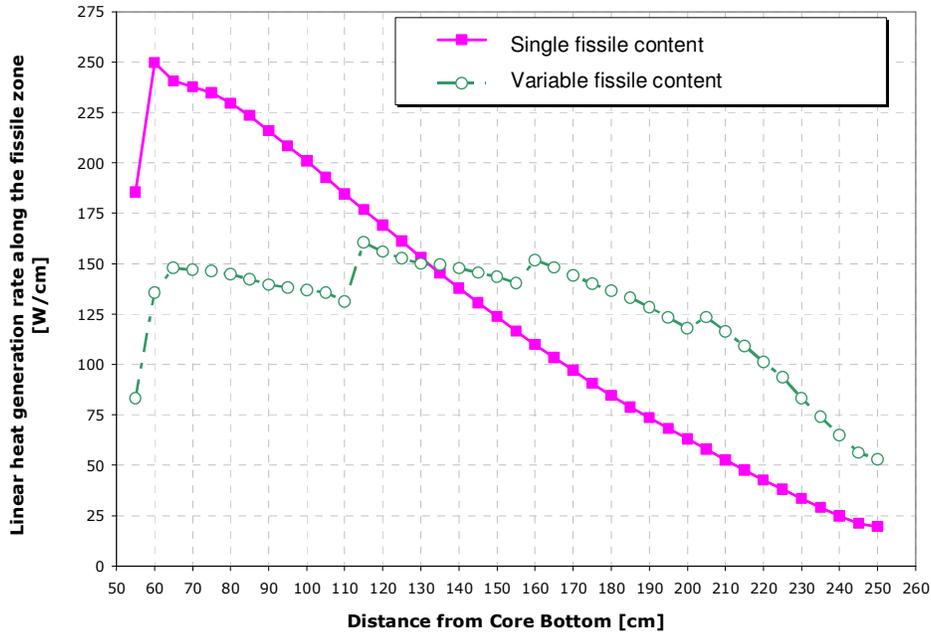

**Figure 8. Linear heat generation rate along the seed section in the Th-RBWR assembly at BOL for uniform and variable fissile content**

The SDM calculations were repeated for the new design with the variable axial enrichment. Table 6 presents the assembly lattice criticality values at BOL for Th-RBWR assembly with variable fissile content.

**Table 6. SDM for Th-RBWR design with variable fissile content**

| Th-RBWR Variable enrichment | ARO | | ARI | | SDM [$\rho_{allCRD}$] |
|---|---|---|---|---|---|
| Case | k-inf | $\Delta$ k-inf | k-inf | $\Delta$ k-inf | |
| HFP | **1.04007** | *0.00026* | **0.85840** | *0.00021* | **0.2035** |
| HZP | **1.41923** | *0.00026* | **1.24333** | *0.00028* | **0.0997** |
| CZP | **1.51899** | *0.00024* | **1.35448** | *0.00026* | **0.0799** |

The table shows that applying variable fissile content in seed section does not notably improve the SDM. However, the maximum linear heat generation rate at BOL was reduced to 160 W/cm, compared to 250 W/cm in the original case indicating considerable potential for power up-rate of the reference design.

**3.5 Use of burnable poisons**

Additional option examined for overcoming the excess reactivity at CZP was the implementation of burnable poison such as Integral Burnable Absorber (IBA), in the form of a thin layer of Gadolinium coating on fuel pellets. The reason for using Gadolinium was its very high thermal absorption cross with relatively low cross section at higher energies. Therefore, it should contribute significantly to the excess reactivity compensation at CZP conditions, while marginally affect the breeding performance at HFP, where the spectrum is much harder and absorption in Gd is small . Figure 9 shows a typical curve of the reactivity versus burnup behavior of burnable poison-loaded BWR fuel assembly.

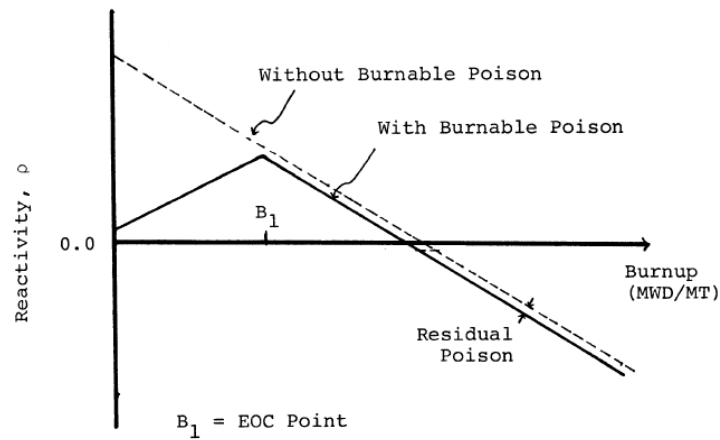

**Figure 9. Generic reactivity versus burnup plot for an assembly with Burnable Poison [Driscoll at al., 1990]**

The number of burnable absorber coated rods in Th-RBWR assembly was adjusted to bring the initial assembly reactivity close to zero at HFP. The assumption was that the IBA-loaded assembly reactivity will behave as presented in Figure 9. It was found that the reference Th-RBWR assembly containing 13 Gadolinium IBA rods (as presented in Figure 10) has reactivity close to zero (i.e. all the excess reactivity is compensated by the IBA rods at HFP).

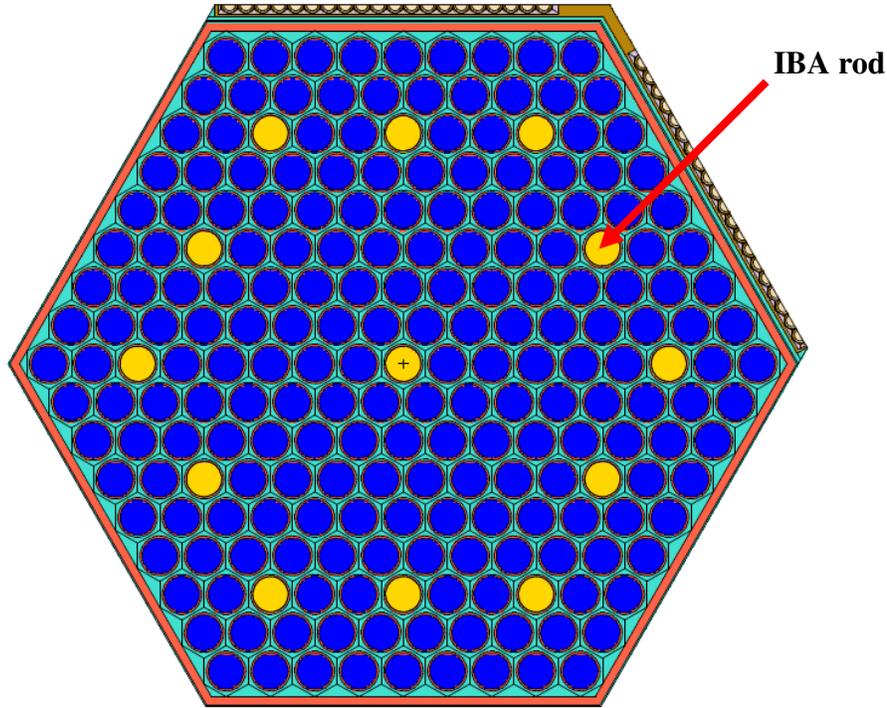

**Figure 10. Th-RBWR fuel assembly with 13 IBA rods and ARI**

Table 7 present the multiplication factor of Th-RBWR assembly with 13 IBA rods at BOL for two conditions:
1. Hot full power with all rods out (HFP-ARO)
2. Cold zero power with all rods in (CZP-ARI)

**Table 7. SDM for Th-RBWR design at BOL with variable fissile content and 13 IBA fuel rods**

| Th-RBWR + Gd IBA | | |
|---|---|---|
| Case | k-inf | *Δ k-inf* |
| HFP-ARO | 0.99327 | *0.00037* |
| CZP-ARI | 1.27950 | *0.00057* |

Table 7 shows that that despite significant contribution of IBA to the reduction of excess reactivity, this is still insufficient to achieve adequate shutdown margin. Moreover, since the neutron spectrum in the seed region at CZP is similar to one at HFP (Fig. 5) and still relatively hard the IBA absorber has practically same effect at both operating conditions. Furthermore, relatively hard spectrum also results in very slow burnup rate of Gd which becomes essentially permanent neutron absorber and therefore, the IBA-loaded assembly reactivity did not behave as presented in Figure 8. In fact, the assembly was subcritical throughout the entire irradiation campaign as illustrated in Figure 11.

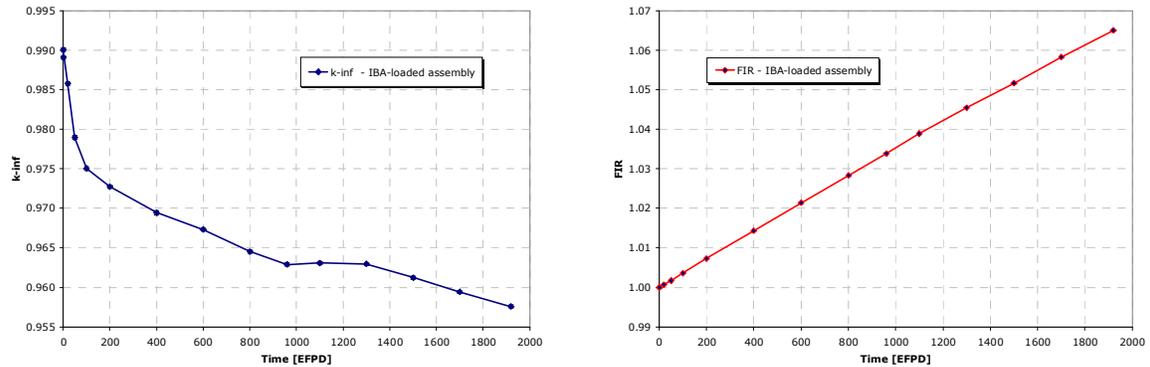

**Figure 11: Assembly $k_{inf}$ (left) and FIR (right) for IBA-loaded assembly**

### 3.6 Effect of fuel assembly size and Control Rod volume

The effect of assembly dimensions on SDM was examined for three assembly layouts shown in Figure 12. In all cases, the assembly flat-to-flat dimensions were modified but the lattice pitch, pin diameters and radial gap distance between the outer surfaces of the fuel pins, assembly can thickness and control rod components thicknesses were preserved. The three examined cases denoted here as: 9×9 layout (217 fuel pins), 6×6 layout (75 fuel pins) and 4×4 layout (37 fuel pins) are also presented in Figure 12. The studied assembly layouts affect the results for a number of important reasons:

- Decreasing the number of pins per assembly increases the neutrons fraction that reaches the assembly periphery thereby increase the worth of the control rod blade. It also effectively increases the number of control rods in the core.
- Changing the amount of heavy metal per assembly, while increasing the relative volume fraction of water filled gaps between the assemblies, increases neutron moderation. Therefore, assembly reactivity is increased, which, in turn, decreases the fissile content and reduces the SDM. This modification, on the other hand, will also negatively affect the conversion ratio and thus overall fissile material balance.

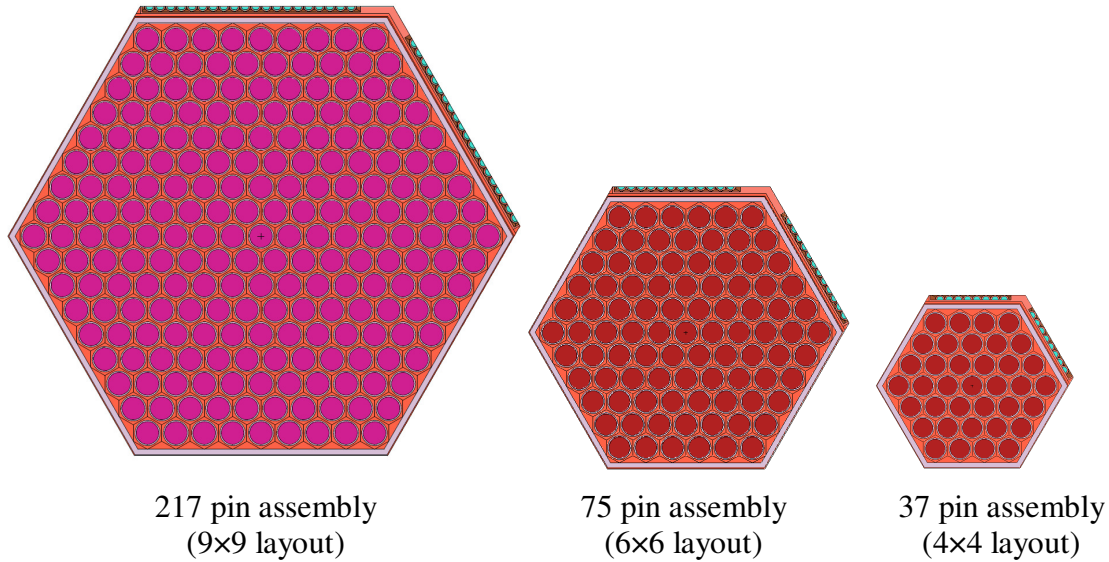

217 pin assembly
(9×9 layout)

75 pin assembly
(6×6 layout)

37 pin assembly
(4×4 layout)

**Figure 12. Radial cross section view of the examined assembly layouts**

Additional option would be to keep the fuel assembly size the same but introduce additional control rods. Theoretically, the assembly could face a control rod on each of the six sides rather than only two. Such an approach was suggested for example for increasing the fuel assembly pitch in conventional BWRs (Yamashita et al., 1991). However, the effect of introducing additional control rods on the shutdown margin is expected to be similar to reducing the assembly size, while keeping the control rods only at two assembly hexagon faces (Fig. 12). In both cases, the relative volume fraction of the control rods in the core would increase and so will the volume fraction of the water gaps. Therefore, this option was not explicitly evaluated in this study.

Table 8 summarizes the SDM calculation results for the examined cases. It was found that decreasing the number of pins per assembly indeed increases the worth of the control rod blade but again, not sufficiently for achieving shutdown margin requirements even for the smallest fuel assembly size considered here.

**Table 8. SDM for three assembly radial layouts of Th-RBWR assembly**

| Case | HFP - ARO | | CZP - ARI | | SDM |
|---|---|---|---|---|---|
| | k-inf | $\Delta$ k-inf | k-inf | $\Delta$ k-inf | [$\rho_{allCRD}$] |
| **Reference 9x9** | **1.05124** | *0.00027* | **1.35171** | *0.00027* | **-0.211** |
| **Variable enr. 9x9** | **1.04007** | *0.00026* | **1.35448** | *0.00026* | **-0.223** |
| **Variable enr. 6x6** | **1.05527** | *0.00020* | **1.29233** | *0.00037* | **-0.174** |
| **Variable enr. 4x4** | **1.07949** | *0.00025* | **1.13310** | *0.00034* | **-0.044** |

## 3.7 Use of Rod Cluster Control Assemblies

Another alternative was to use Rod Cluster Control Assembly (RCCA) similar to those used in PWRs. The RCCA uses several absorber rods that are fastened to a common 'spider' hub (Figure 13). The control rods are inserted into guide thimbles that are distributed throughout the assembly lattice. The absorber material is in the form of extruded AIC rods that are sealed in Stainless Steel tubes. RCCA are primary used in PWRs and inserted from the top.

In BWRs, the upper portion of the core vessel is used for steam conditioning equipment. Therefore, the RCCA can only be inserted from the bottom. Moreover, since the guide thimbles are filled with water at the time the control rods are withdrawn, the hard neutron spectra in Th-RBWR assembly could be affected. In order to minimize the amount of water in the guide thimbles and therefore, reduce the neutron moderation, the RCCA pins should contain water displacing follower on top of the absorber material, as is done in the Y-shape control rod design.

The concept of using RCCAs in BWRs raises a number of important questions with regards to their structural design because in BWR, the control rods could be subjected to very different mechanical loads and operating conditions. We do not attempt to address these essential issues in this study but rather concentrate first purely on the neutronic performance of the concept. Structural design issues will need to be addressed in the future should the concept is proved to be feasible from the neutronic performance point of view.

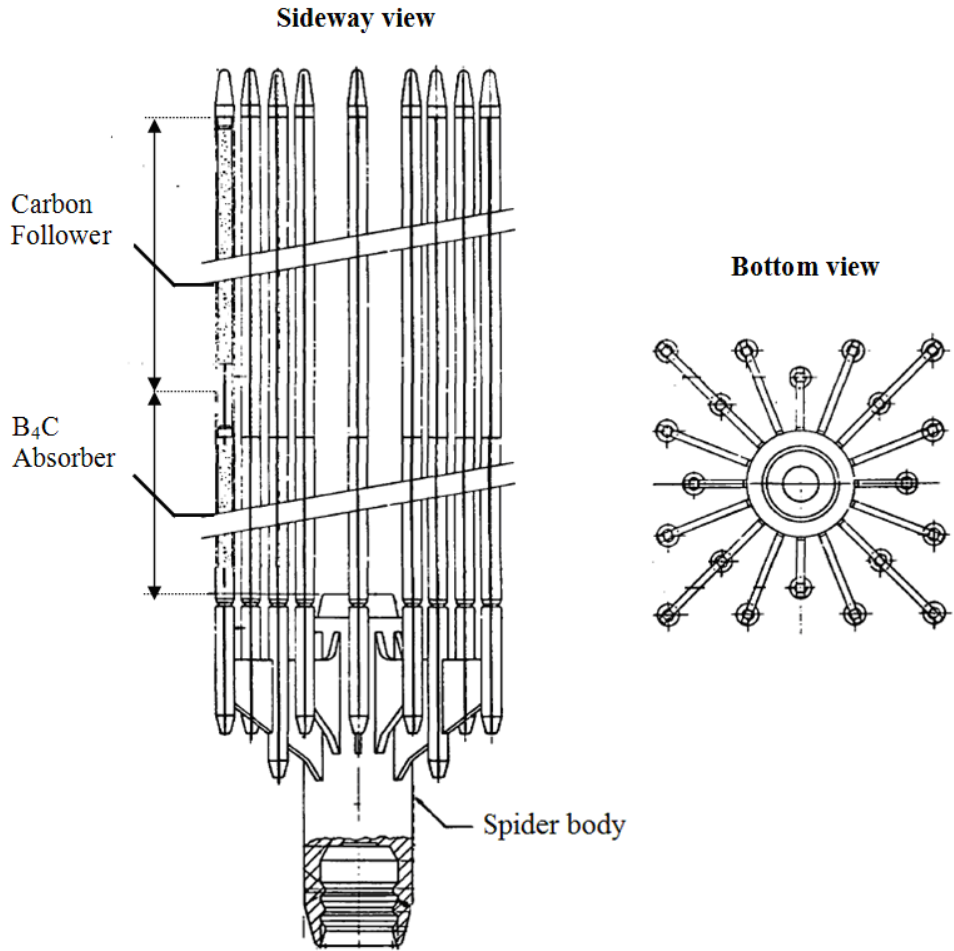

**Figure 13. Radial view of RCCA spider**

Figure 14 presents one of the studied Th-RBWR assembly designs with RCCA containing 19 control rods (up to 25 control rods per assembly were considered). Table 9 presents the dimensions and composition of the guide thimble and the control rod. The dimensions were adopted from a typical PWR design and may need to be changed due to mechanical design constraints. The absorber material used was $B_4C$ with 90% $^{10}B$ enrichment.

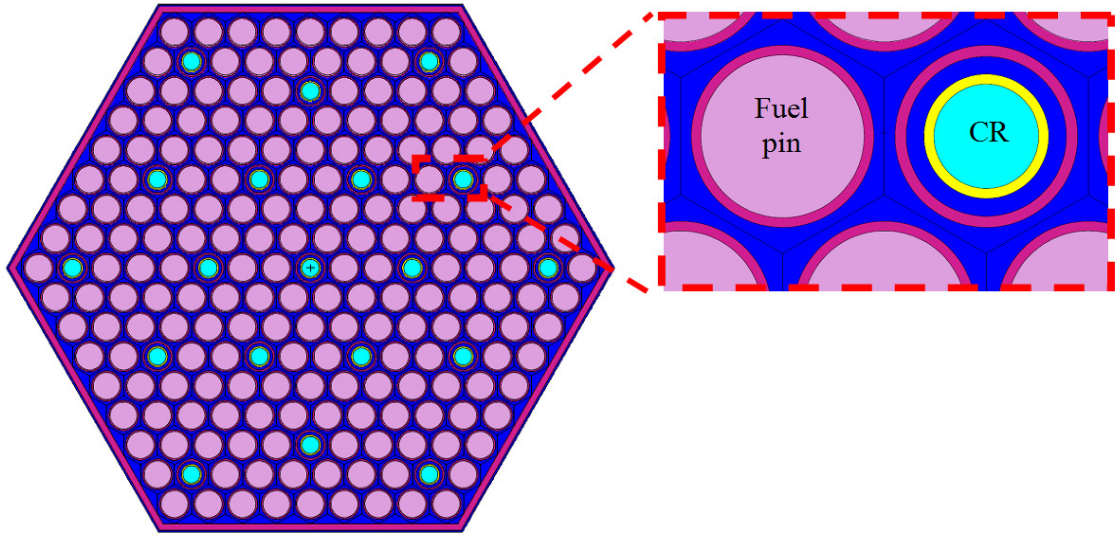

**Figure 14. Radial view of Th-RBWR assembly with RCCA spider containing 19 control rods**

The outer diameter of the Zircaloy guide thimble was set to be equal to the outer diameter of the fuel pin cladding in order to preserve the lattice geometry. The Y-shape control rod used in the reference Th-RBWR design was removed from the inter assembly gap and the flat-to-flat assembly dimensions (assembly pitch including the water gap) were adjusted accordingly to 19.52 cm.

**Table 9. RCCA radial dimensions**

| Parameter | Value | Units |
|---|---|---|
| Absorber / Follower diameter | 0.6478 | [cm] |
| Stainless Steel tube thickness | 0.0620 | [cm] |
| Inner diameter of the Zircaloy guide thimble | 0.9918 | [cm] |
| Outer diameter of the Zircaloy guide thimble | 1.1298 | [cm] |

Four cases were examined: Case A – contained 7 control rods (210 fuel pins), Case B – contained 13 control rods (204 fuel pins), Case C – contained 19 control rods (198 fuel pins), and Case D– contained 25 control rods (192 fuel pins) as presented in Figure 15. The position and the number of the control rods was chosen such that $^1/_6$ symmetry of the assembly was preserved. The fissile content in the seed zone was again divided into five regions containing variable fissile contents (3.5-7.5$^w/_o$ of $^{233}$U) whereas the average fissile content was fixed at 6.0025$^w/_o$ of $^{233}$U.

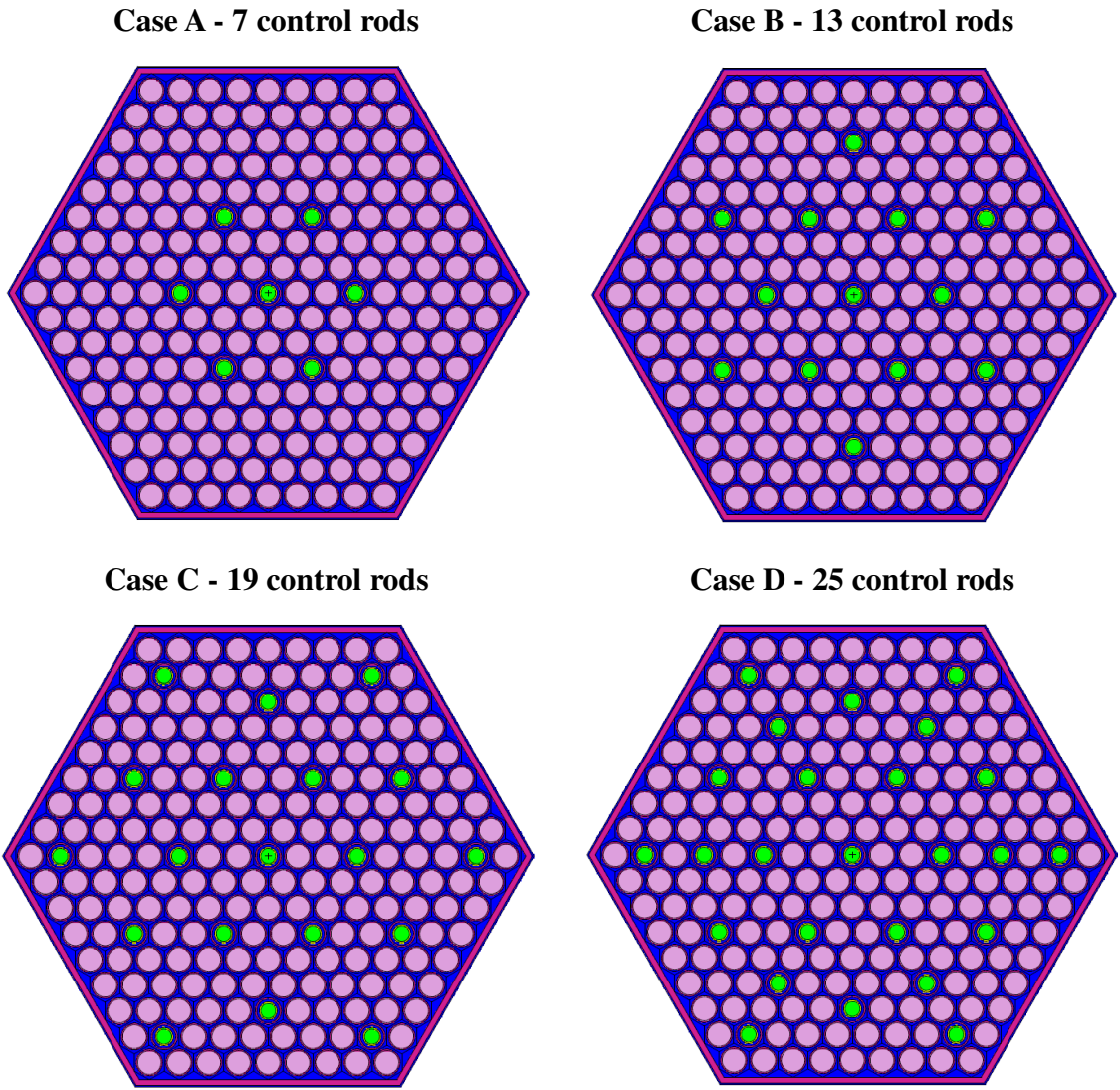

**Figure 15. Radial view of Th-RBWR assemblies with RCCA spider**

Introduction of RCCA instead of the Y-shaped control blade allows for a larger absorber surface area which faces the fuel as well as increase in the amount of absorber material. As a result, the RCCA is more efficient than the Y-shape control rod. Table 10 summarizes the assembly multiplication factor for Cases A-D at BOL for cold zero power conditions with ARI.

**Table 10. Criticality of Th-RBWR assembly lattice with RCCA**

| Case | BOL – CZP – ARI | |
|---|---|---|
| | k-inf | Δ k-inf |
| 7 control rods (Case A) | **1.33828** | *0.00025* |
| 13 control rods (Case B) | **1.18538** | *0.00025* |
| 19 control rods (Case C) | **1.03358** | *0.00022* |
| 25 control rods (Case D) | **0.92562** | *0.00024* |

Table 10 presents values obtained with fissile content representative of the reference design (i.e. 6.025$^w/_o$). Addition of water in the guide tubes slightly changes the spectrum. Therefore, this reference fissile content needs to be slightly readjusted (reduced) for each case to sustain desired cycle length. This will also reduce the $k_{inf}$ values presented in Table 10. It was found that Case C, which contained 19 control rods, can ensure adequate shutdown margin as in Hitachi RBWR design [Feng et al., 2011] once the leakage and partially burnt batches effects are taken into account.

Additional concern of the Th-RBWR assemblies with RCCA is that the thermal power of the assembly may need to be further reduced compared to the reference Th-RBWR design due to the smaller number of fuel pins. Furthermore, the additional water in the guide thimbles will cause undesirable neutron moderation and further reduce the FIR. Therefore, the neutronic and thermal-hydraulic aspects of Case C design are further studied in the following section.

## 4. Neutronic and thermal-hydraulic performance of an assembly with RCCA

This section examines the effect of the RCCA on the neutronic and thermal hydraulic performance of Th-RBWR. The main parameters of interest are:

- The effect the RCCA on FIR performance.
- The achievable thermal core output relative to ABWR, imposing the limits on Critical power ratio (CPR) and peak fuel temperature.

In order to address these issues, two cases were analyzed; one case preserves the power to mass flow rate ratio used in the reference Th-RBWR design. The other case represents an assembly with higher power, to match the ABWR's thermal core output of 3973 MW$_{th}$. Results obtained for the two cases were compared to the reference Th-RBWR assembly and to the case identical to reference one but with axially varied fissile content. In each case, the total fissile content per assembly was iteratively adjusted to obtain a fuel cycle ($T_{cycle}$) of 24 month assuming a three batch core fuel management and applying LRM assumptions with 0.4% allowed for radial neutron leakage.

The case descriptions are summarized below:

**Case 1:** Reference Th-RBWR assembly containing 217 fuel rods with axially uniform fissile content

**Case 2:** Th-RBWR assembly containing 217 fuel rods with variable fissile content

**Case 3:** Th-RBWR assembly containing 198 fuel rods and 19 control rods, with variable fissile content. The power to mass flow rate ratio is identical to Case 2

**Case 4:** Th-RBWR assembly containing 198 fuel rods and 19 control rods, with variable fissile content. The assembly power was increased to match the ABWR core power

Table 11 presents the main thermo-hydraulic parameters of the examined cases.

**Table 11. Main thermo-hydraulic parameters of Cases 1-4**

| Case<br>Parameter | 1&2 | 3 | 4 |
|---|---|---|---|
| Assembly thermal output [MW$_{th}$] | 5.51 | 5.025 | 7.9 |
| Assembly mass flow rate [kg/sec] | 6.64 | 6.065 | 11.5 |
| Core thermal output [MW$_{th}$] [†] | 2781 | 2537 | 3989 |
| Average fissile content [$^w/_o$] | 6.50[††] | 5.27 | 5.70 |
| Assembly heated perimeter [cm] | 770.21 | 702.77 | |
| Assembly flow area [cm$^2$] | 87.74 | | |
| Assembly wetted perimeter [cm] | 847.77 | | |

[†] Assuming that Th-RBWR core has 505 hexagonal assemblies [Shaposhnik et al. 2013]

[††] In case of variable fissile content (Case 2), the assembly average fissile content was 6.0025$^w/_o$

Figures 16 and 17 present the assembly $k_{inf}$ and FIR as a function of time for Cases 1-4.

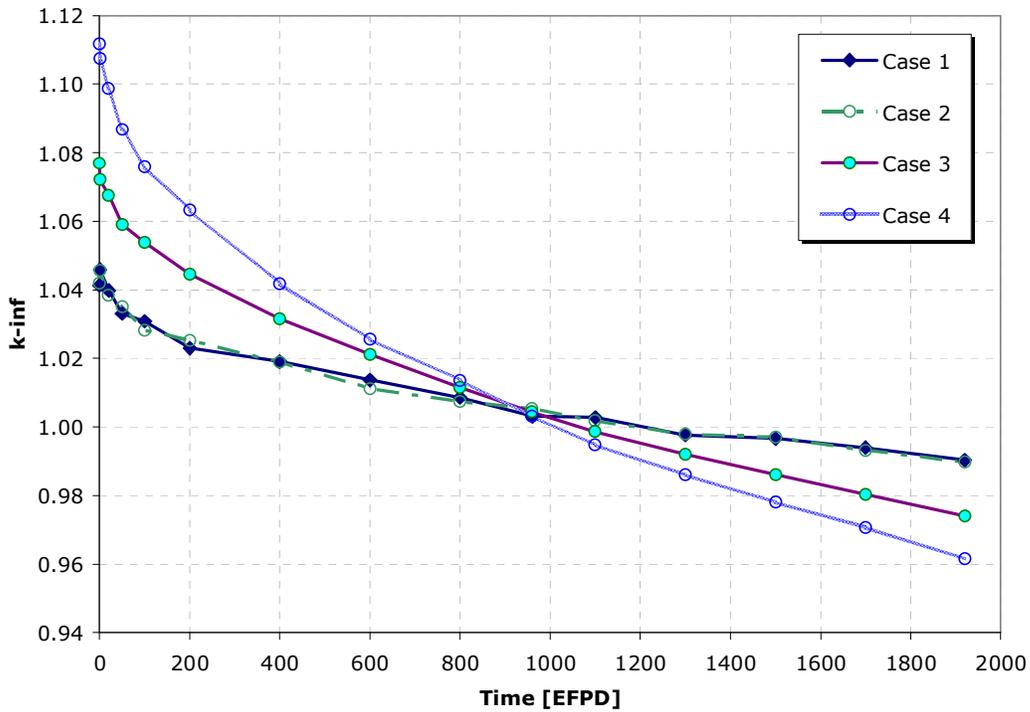

**Figure 16: $k_{inf}$ for Cases 1-4**

In Figure 16, the excess reactivity at BOL is higher for the assemblies containing smaller number of fuel rods (Case 3 and 4) due to the softer spectrum resulting from additional moderation in the water present in the guide thimbles. As a result, the FIR at fuel discharge (Figure 17) is reduced to 1.045 in Case 3 and to 1.021 in Case 4 respectively compared to 1.074 in the reference case (Case 1). Furthermore, in Case 4 the power to mass flow rate ratio was also reduced compared to the other cases, which results in lower core average void fraction, also softening the spectrum and reducing the FIR (Figure 17). Figure 18 presents the axial void distributions for Case 1-4 at BOL.

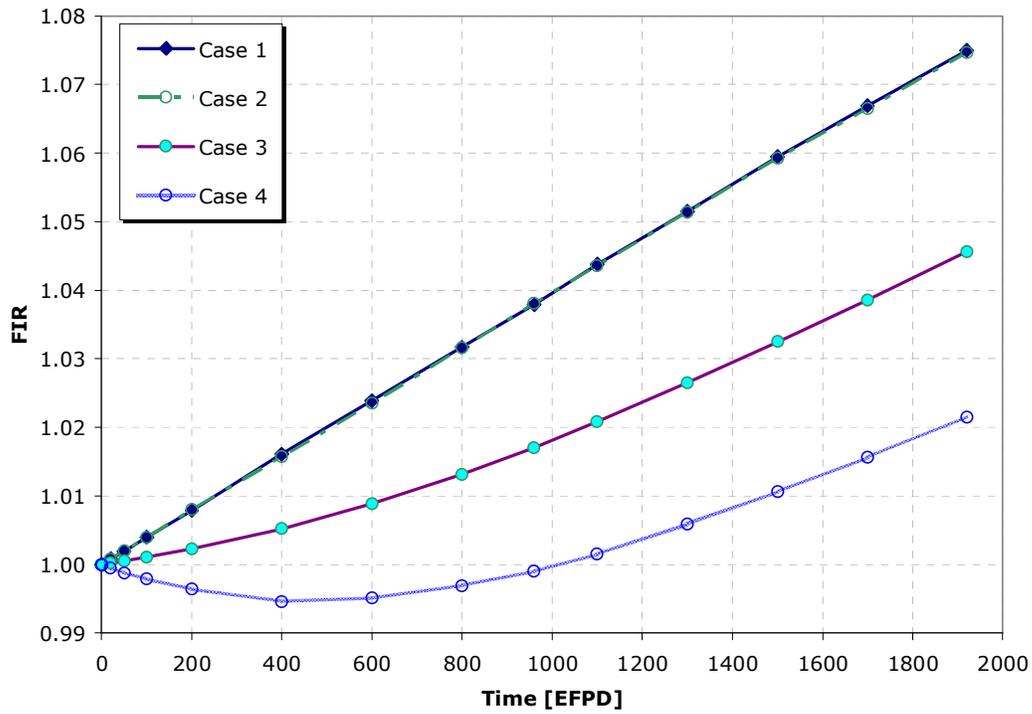

**Figure 17: FIR for Cases 1-4**

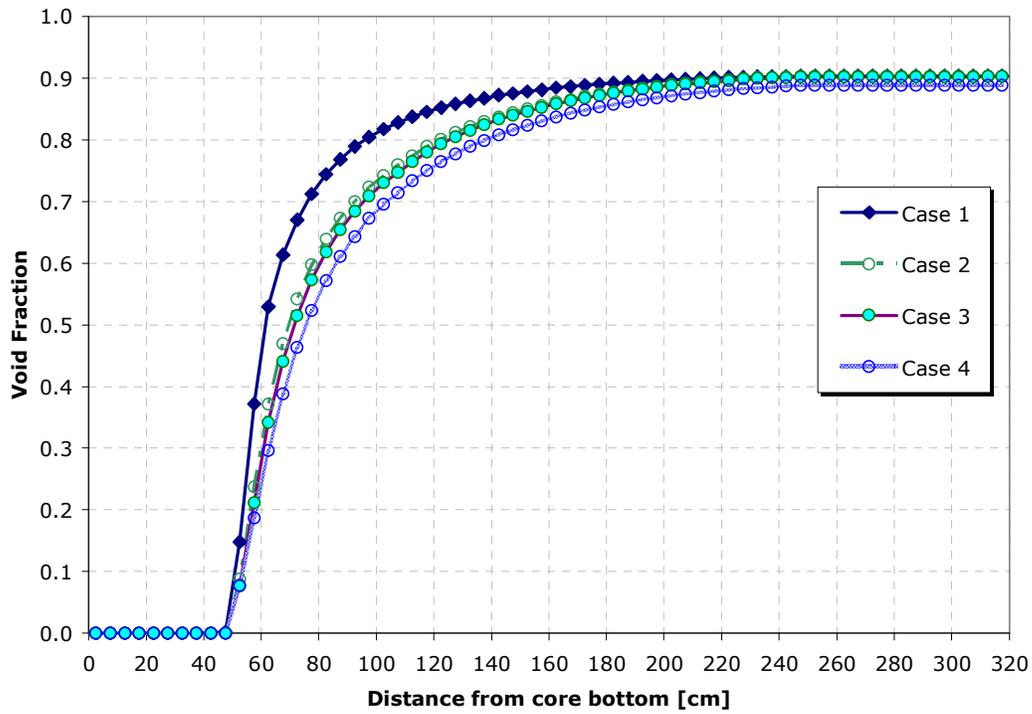

**Figure 18: Axial void distribution in Cases 1-4 at BOL**

The mass flow rate in Case 4 was chosen as a compromise between FIR and the Minimum Critical Power Ratio (MCPR). The MCPR in Th-RBWR assemblies is quite sensitive to the power to flow ratio [Shaposhnik et al. 2013]. Decreasing the power to flow ratio increases the CPR and allows uprating the assembly thermal power. However, at the same time, it reduces the assembly average void fraction, resulting in a softer spectrum and lower FIR.

MCPR is evaluated using Liu et al. (2007) CPR correlation, which is based on the quality-boiling length relation. It considers both local and channel dryout mechanisms and accounts for reduced moderation (tight pitch) conditions. The Liu et al. (2007) correlation was implemented in the THERMO module for predicting the CPR of the studied assembly designs. Figure 19 presents the actual equilibrium vapor quality ($x_e$) and critical vapor quality ($x_{Crit}$) profiles for Case 1 and 4.

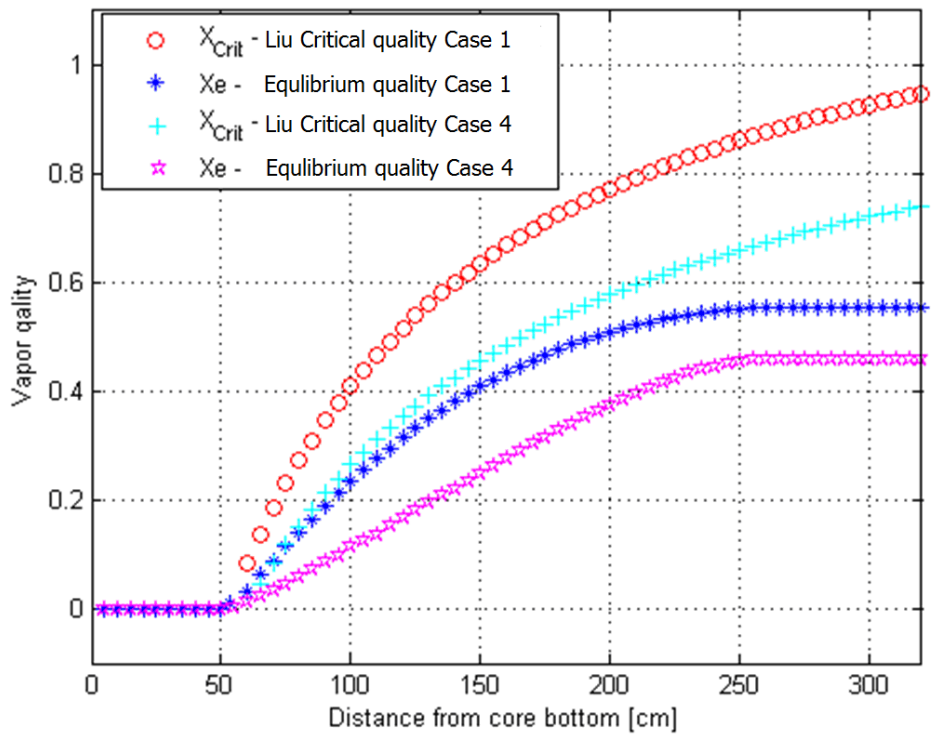

**Figure 19: Equilibrium vapor quality and critical quality profile at BOL for Case 1 and 4**

The most restrictive time, with respect to the CPR, occurs at BOL for all designs. This is because the axial power peaking factor at BOL is the highest. It decreases with irradiation due to the gradual buildup of fissile material in the blankets and the depletion of fissile material in the seed. The MCPR at BOL for the reference Th-RBWR design (Case 1) was 1.50 and for Case 4 is 1.44 at BOL, 1.47 at MOL and 1.50 at EOL. This was deemed to be acceptable assuming the assembly radial peaking factors in the fresh fuel batch can be kept below 1.3 through the loading pattern optimization.

Figure 20 presents the axial void profile along the Case 4 assembly for BOL, MOL and EOL. The coolant enters the assembly as nearly saturated liquid. The blanket region at BOL does not experience any boiling, since, at BOL, the blanket practically does not produce power. As the assembly undergoes depletion, the blankets start to accumulate fissile material ($^{233}$U) and to produce power resulting in a gradual shifting of the void profile with burnup (Figure 21). The performed analyses underscore the importance of including thermal-hydraulic feedback in design studies.

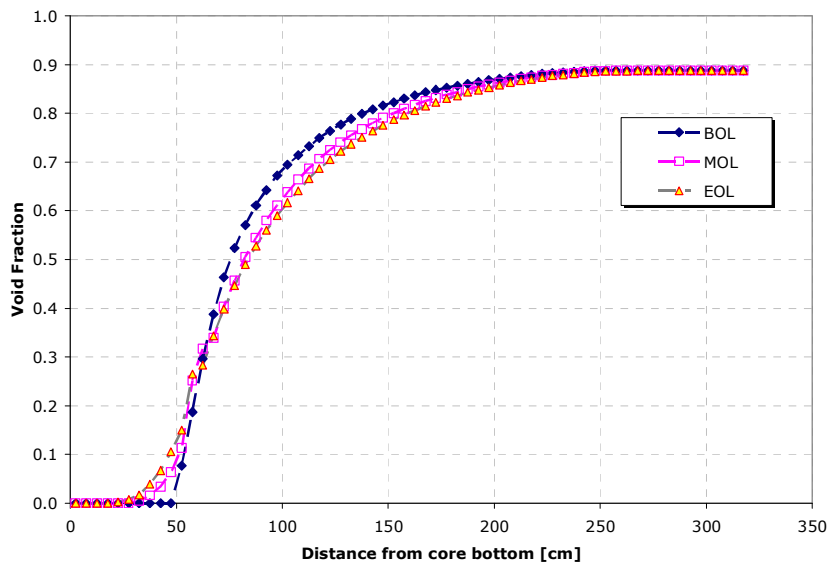

**Figure 20: Axial void profile in an average fuel assembly of Case 4 at BOL, MOL and EOL**

Figure 21 presents the linear heat generation rate and fuel centerline temperature along the fuel assembly at BOL, MOL and EOL. The BGCore model for calculating the centerline and average fuel temperatures accounts for temperature dependence of the fuel thermal conductivity for $ThO_2$-$UO_2$ fuels as suggested in IAEA (2006) report. The maximum linear heat generation rate was 260 W/cm at the second axial region of the seed with variable fissile content (112 cm from core bottom) at BOL. The corresponding maximum fuel centerline temperature was 846 $^{o}$C which is comfortably below the melting point even with allowance for anticipated transients.

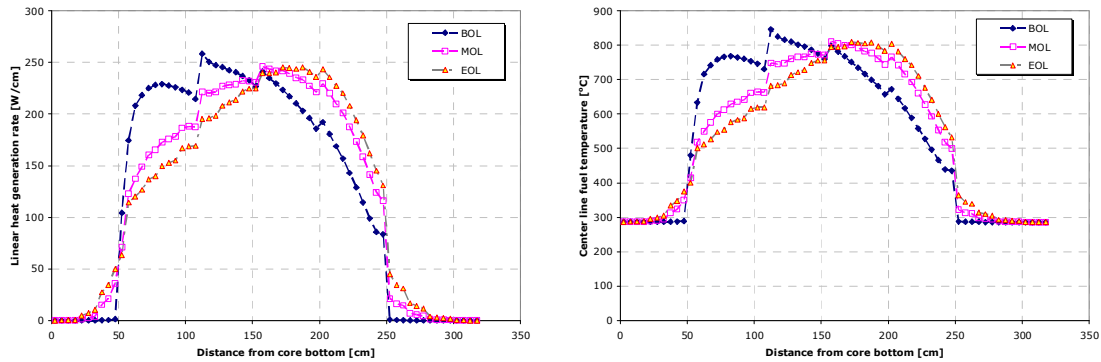

**Figure 21: Linear heat generation rate (left) and centerline fuel temperature (right) along Case 4 an average fuel assembly at BOL, MOL and EOL**

The SDM for Case 4 was evaluated at three points during the fuel irradiation campaign: BOL, MOL and EOL. For each point, SDM was calculated for two conditions: HFP-ARO and Xenon free CZP-ARI, the results are given in Table 12.

**Table 12. SDM for Case 4 design BOL MOL and EOL**

| Case | HFP-ARO | | CZP-ARI | | SDM |
|---|---|---|---|---|---|
| | k-inf | $\Delta$ k-inf | k-inf | $\Delta$ k-inf | [$\rho_{allCRD}$] |
| BOL | **1.11175** | *0.00017* | **1.01840** | *0.00025* | **0.082** |
| MOL | **1.00311** | *0.00016* | **0.90879** | *0.00025* | **0.103** |
| EOL | **0.96158** | *0.00016* | **0.85935** | *0.00022* | **0.124** |

The fuel assembly reactivity behavior, as presented in Figure 15, is similar to that of conventional fuel assembly with enriched uranium fuel (monotonically decreasing, nearly linear curves). It is known from experience that for such reactivity curve shapes, the traditional approach for estimating the core average parameters based on the Linear Reactivity Model [Driscoll et al., 1990] yields reasonably accurate results. In the examined case, the core operates in a three-batch in-core fuel management with three fixed fuel cycle intervals and assumed to have equal power sharing between the batches.

Therefore, for a three-batch core at steady state, the core average reactivity can be estimated as arithmetic average of individual batch reactivities. Table 13 presents the core $k_{inf}$ calculated using the described LRM assumptions and SDM (Xenon free) at BOC and EOC for the Case 4 design.

Table 13. Core $k_{inf}$ and SDM for Case 4 design

|  | Core k-inf HFP-ARO | Core k-inf CZP-ARI | Core SDM [$\rho_{allCRD}$] |
|---|---|---|---|
| BOC | 1.0411 | 0.9443 | 0.099 |
| EOC | 1.0038 | 0.9040 | 0.110 |

Table 13 shows that shutdown margin can be comfortably met for Th-RBWR Case 4 assembly design. This design contains 198 fuel rods and 19 control rods that are arranged in common to PWRs 'spider' hub, match the ABWR core power rating while achieving FIR above unity at EOL.

**CONCLUSIONS**

Several design options were explored for a BWR based on axially heterogeneous fuel assembly, operating in a self-sustainable Th-$^{233}$U fuel cycle in order to meet the shutdown margin requirements. The analysis was restricted to an assembly with a single axial fissile zone sandwiched between two fertile blanket zones. The study covered several design parameters including absorber material, burnable poison loading and control rods layout.

The first phase of the study examined the possibility of achieving adequate shutdown margin in Th-RBWR design using the Y-shape control rod as in the reference RBWR-AC assemblies. In order to account for spatial effects, the Y-shape control rod had to be modeled in details. The RBWR-AC design was used as a reference case for SDM calculations. The assembly radial dimension as well as the control rods design of Th-RBWR was identical to the RBWR-AC design. Despite similar reactivity at BOL-HFP-ARO conditions in both designs, at BOL-CZP-ARO the Th-RBWR assembly had substantially higher excess reactivity of about 50,000 pcm compared to about 7,000 pcm in the RBWR-AC design. This difference is partly due to the axial configuration of the fissile regions and partly due to the large magnitude of negative VC and DC. Therefore, the Y-shape control rod design of the RBWR-AC was found to be unsuitable for the Th-RBWR design since it does not allow achieving adequate shutdown margin.

In the second phase, the SDM for various reactivity control materials used in Y-shape control rods were analyzed. This included boron, hafnium, gadolinium, and a mix of silver, indium and cadmium isotopes. The main conclusion was that $B_4C$ with high $^{10}B$ enrichment has the highest reactivity worth in Th-RBWR assembly design.

In the third phase, we examined the possibility of reducing the leakage from the seed zone to the blanket zones and thus reducing the peak linear heat generation rate. This was achieved by re-distributing the fuel fissile content (variable fissile content) in the seed zone. Based on the fuel assembly analysis, it was concluded that implementation of variable fissile content can reduce the initial reactivity and the peak linear heat generation rate without affecting the FIR. Nevertheless, the shutdown margin was still insufficient.

In order to suppress the assembly excess reactivity at CZP, the use of burnable poisons (IBA) was examined. The IBA loaded assembly contained 13 Gadolinium IBA fuel pins, chosen because if its higher absorption cross section in the thermal region, thus being efficient absorber material at CZP conditions, while at HFP its absorption effect is small due to the harder RBWR neutron spectrum. As a result the effect on FIR is also small. Despite significant contribution of the IBA to reduction of excess reactivity at CZP, it was still insufficient to meet the shutdown margin requirements.

Another option considered for achieving SDM was reducing the number of pins per assembly in order to increase the fraction of neutrons that reach the assembly periphery and get absorbed by the control rod. Three layouts were examined, in which the assembly flat-to-flat dimensions were modified but preserving the lattice pitch, pin diameters and radial gap distance between the outer surfaces of the fuel pins, assembly can and control rods. As expected, decreasing the number of pins per assembly increases the worth of the control rods blade but, again, less than required for achieving adequate shutdown margin.

Finally the possibility of replacing the Y-shape control rods with a Rod Cluster Control Assembly (RCCA) 'spider' hub with absorber rods was studied. It was found that 19 control rods per assembly are sufficient to meet the shutdown margin requirements.

Moreover, due to implementation of variable fissile content, the assembly power could be further increased within the same thermal limits. However, in order to meet CPR margin, the assembly mass flow rate had to be increased as well, which in turn reduced the assembly average void fraction, softened the spectrum and reduced FIR. Nevertheless, FIR at the end of the fuel irradiation campaign was still above unity for the design that matches the standard ABWR core power.

In summary, an assembly design relying on axially heterogeneous fuel zoning, which can achieve FIR above unity at the end of fuel irradiation campaign and achieve adequate shutdown margin was identified. It features hexagonal pins and assembly lattice, with 198 fuel pins and 19 control rods per assembly, a single axial seed section of 200 cm high sandwiched between two fertile blanket zones. The core operates in a 24 month cycle with three batch reload. The Fissile Inventory Ration (FIR) for this design at the fuel discharge is 1.02 and the thermal core output can potentially match the ABWR power of 3973 MW$_{th}$.